# Results of UBV Photoelectric Observations of eclipsing binary RY Sct.

**M.I.Kumsiashvili, K.B.Chargeishvili, E.B.Janiashvili**
Ilia State university E. K. Kharadze Abastumani Astrophysical Observatory, Georgia;
e-mail: kumsiashvili@genao.org, ketichargeishvili@iliauni.edu.ge,
edik_var@yahoo.com

## Abstract

Results of the three-colour photoelectric observation of the close binary system RY Sct , obtained at the Abastumani Astrophysical observatory, are presented.

## Introduction

In the 1980'ies, by the decision of one international forum Abastumani Observatory was charged with coordination of cooperated research of early type unusual eclipsing binary system RY Sct. Thereby, on the basis of coordinated program an extensive research of these unique astrophysical objects was conducted at Abastumani Astrophysical Observatory during years (coordinator M. Kumsiashvili). Three color and six color photoelectric observations of this object were carried out at Abastumani Observatory during the period of 1972-1990 by M. Kumsiahsivli. Having entered the international scene we obtained spectral and polarimetric observations with the help of a well-known European researcher R. West (ESO). We included into the program well-known specialists of this sphere from the former Soviet Union, Europe and America. The most different result, obtained for this object, was the identification of spectral lines of secondary component and, as a result, receiving mass ratio equal to 3.3 instead of the value 1.25, recognized earlier. This has substantially changed our conception of this system. The results of coordinated studies were published in the Bulletin of Abastumani Astrophysical Observatory No.58 (Kumsiashvili, a), b); Skul'skij; Babaev; Karetnikov; Cherepashchuk; Zakirov; 1985) under the editorship of Kumsiashvili. The results of complex investigation were presented by Kumsiashvili in 1988 at IAU International Conference "Algols" in Canada (Kumsiashvili, 1989). The results were also published in Russian Astronomical Journal (Antokhina and Kumsiashvili, 1999), as well as in journal "Revista Mexicana de Astromomia y Astrofisica" (Sahade et. al. 2002).

We had almost finished works, dedicated to the investigation of this object, when the new works of American investigators appeared in literature on this object with new results (Smith, Gehrz, 1999-2002). They basically concerned the nebula, surrounding the system, and contained an indication regarding the fact that at least one component of RY Sct must be a LBV-type star. In their opinion it was this star that about 120-200 years ago experienced an LBV-type outburst, which was the basis for the origin of a young nebula around the system with an unusual geometry. After that information we, the group of Georgian scientists, decided to return to our old photoelectric material, which we kept in archive, and to study the character of variability related to physical processes in order to confirm or reject the Americans' assumptions. We contacted our American colleagues and our collaboration was executed in the form of a joint grant (**GEP1-3333-TB-03**). The project began in December 2004 and was completed in May 2006. We have obtained significant results. Specifically, on the basis of available photoelectric material microvariations were revealed, which are typical for LBV stars after outburst. In addition, observations of the first maximum of light curve show a long-period variability, which must be related to the physical processes in the component eclipsing in the main minimum. Proceeding from the above stated, with our data we actually confirmed the Americans' assumption that RY Sct is as yet the only binary system, which may contain as a component the LBV-type star, as a result of explosion of which about 120-200 years ago a nebula emerged around the system. In our opinion further observations of this unique astrophysical object will allow us to solve many other interesting problems. Specifically, in conditions of the presence of sporadic mass ejections and intensive gas flows variability of the period of system may be expected. We were not able to solve this problem with the available data. Further accurate photoelectric observations exactly in the depths of moments of minimum are required. Besides, it is interesting to patrol this system by observations in order to register possible further outbursts.

As it follows from the above stated, currently it is very interesting also to study the connection between massive stars and the environment surrounding stars, as well as to learn about the mechanisms leading towards collapse and, as a result, the ejection from a massive star. Of no less interest is the study of connection between the precursor stars, the mechanism of outburst and supernova remnant (SNR) (Dubner 2008).

The problems, unresolved till now, are standing in the way of understanding these issues. Specifically, we do not know about physical processes leading from massive star to the explosion of SN. Clarification of these issues is really of critical importance for linking the supernova predecessor to its remnants.

Indeed, as has been quite reliably established by now, the famous SN 1987 A is the result of ejection from a blue supergiant (BSG). A special problem therefore is to study the formation of a ring-shaped nebula around SN 1987 A and compare it with similar ring-shaped nebulae around galactic blue supergiants. As indicated by Smith (2007, 2008), for instance, the B1.5 supergiant Sher 25 HST image in NGC 3603 cluster shows a noticeable equatorial ring with the same radius as we have for SN 1987 A. Its LBV status is very interesting for, as we know, such stars are characterized by the episodes of outbursts with high losses of mass (Smith and Owocki, 2006), often surrounded by bipolar nebulae. Based on the properties of nebula he asserts that the nebulae might have been ejected from LBV (Smith 2007).

If we consider LBV as the predecessor of SN, it will lead us to further uncertainty regarding the interpretation of LBV-type nonstable phenomena. The case is that according to the existing supernovae flare model the above mentioned supernova prior to flare must have had the structure of a red supergiant, while, as was reliably established, SN 1987 A prior to flare is a blue supergiant. Naturally, as a result theoreticians were confronted with many dilemmas, the more so that the supernova belongs to the second type of supernovae and is the first star of such type to have erupted in a Magellan clouds-type irregular galaxy. Proceeding from the above stated it can be easily understood how important the detailed study of luminous blue stars is, the more so if they represent the components of binary systems.

In such situation it should be noted that the star RY Sct is at a special stage of its evolution. It is a predecessor of WR+OB system. We can draw such a conclusion first of all as a result of conducted spectral investigations of this object, which show that the atmosphere of primary component is rich in helium. Specifically, as a result of exchange of matter this star has developed many properties of helium stars.

The LBV-type variable has a spectrum, very similar to the one of the famous star P Cyg. In other words, according to the spectrum we can see that the loss of matter occurs out of the star. Besides, it is presumed that the light curve of S Dor might show the behavior of a long period eclipsing-variable. It is highly probable that the LBV-type variability may be revealed based on multicolored photometric observations. This is stipulated by the fact that it is characterized by nonstability of several types, such as, for example, "SD-eruption", variability caused by pulsation (SD-phases), which may be occurring with the interval of several years, microvariations with the interval of several days or months. Such phenomena may be detected through the analysis of light curves plotted based on the accurate multicolored photometric observations.

As it follows from the above review with our supposition old photoelectric material, which we kept in archive of Abastumani astrophysical observatory may be used in future repeated for the resolving many interesting problems about RY Sct. That is why we decided make public our individual photoelectric observations in UBV according to nights, which was not published completely till here.

## Observations

Three-color photoelectric observations were carrying out at the abastumani observatory on Mount Kanobili. These observations were made with 0.48 m reflector AZT-14A and AFM-6 photoelectric photometer. An FEU-78 electron multiplier as the detector of light and standard glass Schott filter for defining the UBV photoelectric system were used. The method of pulse counting was applied. The observations were made in 1972-1985. BD-$12^o$5049 was used as the comparison and BD-$12^o$5049 served as the reference star.

The observations made at Mount Kanobili in the UBV system cover the whole light curve and are so far the best and most extensive set of the observations of this star.

In each color separately more then 1000 individual observations were carried out. The fact that in 1972, 1973 observations of RY Sct (5 nights altogether) were done with a self-recorder and beginning from 1975 a photon counter was used, is noteworthy. Accordingly, shifting of observations was observed in these years. In presented observations this fact was taken into account. Altogether 106 observational nights are at our disposal.

Orbital phases were computed using the Ciatti et al. (1980) ephemeris formulae.

$$\text{Min I} = 2443342.42 + 11^d.12471E \quad (1)$$

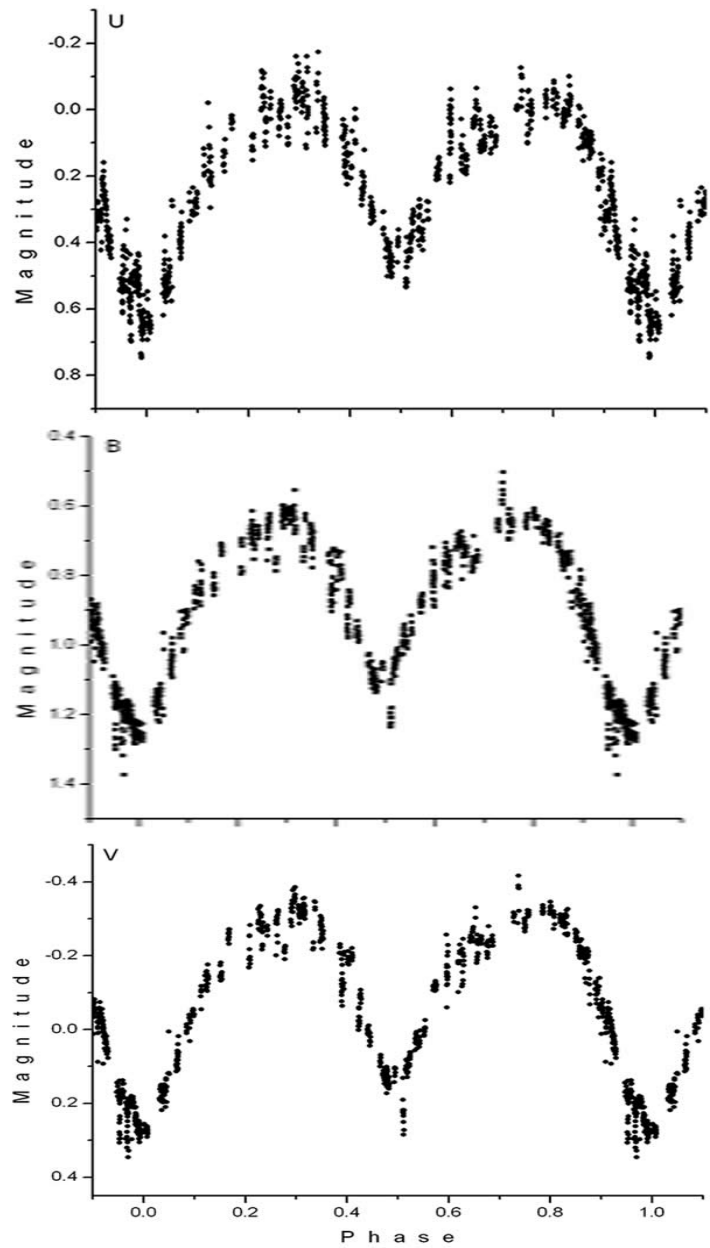

Fig.

The individual differential observations, corrected for differential extinction are given in figure and tabulated in table. Here the first column contains the heliocentric Julian date, the second column contains the phase computed with the ephemeris (1); and the last three column shows the magnitude differences between the comparison and variable stars in yellow, blue and ultraviolet rays, respectively.

References:

Antokhina E.A., Kumsiashvili M.I., 1999, Astronomy Letters, v. **25**, p.662.
Babaev, M.B., 1985, Bull.of Abast. Astroph. Obs., No **58**, p. 105.
Cherepashchuk, A.M., 1985, Bull.of Abast. Astroph. Obs., No **58**, p. 113.
Ciatti, F.; Mammano, A.; Margoni, R.; Milano, L.; Vittone, A.; Strazzulla, G., 1980, Astronomy and Astrophysics Supplement Series, vol. **41**, p. 143-149.
Gehrz, Robert D.; Smith, Nathan; Jones, Barbara; Puetter, Richard; Yahil, Amos., 2001, ApJ, v **559**, p. 395.
Grundstrom, E.; Gies, D.; Hillwig, T.; McSwain, M. ; Smith, N.; Gehrz, R.; et.al., 2007, ApJ, v.**667**, p. 505.
Dubner G., 2008, Rev. Mex. de Astron. Astrof. , ser. de confer. , v **33**, p 148.
Smith, N.; 2007., ASCP, p.200.
Smith, N., 2007., AJ, v.**133,** p.1034.
Smith, N., 2008., Rev. Mex. de Astron. Astrof. , ser. de confer. , v **33**, p 154.
Smith, N.; Owocki,S.P., 2006, ApJ, v.**645**, 245.
Karetnikov, V.G., 1985, Bull.of Abast. Astroph. Obs., No **58**, p. 111.
Kumsiashvili, M.I., 1989, Sp.Sc.Rev., v **50**, No ½, p.350.
Kumsiashvili, M.I., a). 1985, Bull.of Abast. Astroph. Obs., No **58**, p. 61.
                b). 1985, Bull.of Abast. Astroph. Obs., No **58**, p. 93.
Sahade J., West R.M., Skul'skii M.Yu., Rev. Mex. de Astron. Astroph., 2002, v **38**, p 259.
Skul'skij, M. Yu., 1985, Bull.of Abast. Astroph. Obs., No **58**, p. 101.
Smith, Nathan; Gehrz, Robert D.; Humphreys, Roberta M.; Davidson, Kris; Jones, Terry J.; Krautter, Joachim, AJ, 1999. v **118**, p 960.
Smith N., Gehrz R.D., Goss W.M., AJ, 2001, v **122**, p. 2700.
Smith, Nathan; Gehrz, Robert D.; Stahl, Otmar; Balick, Bruce; Kaufer, Andreas., ApJ, 2002, v **578**, p.464.
Zakirov, M.M., 1985, Bull.of Abast. Astroph. Obs., No **58**, p. 425.

Table

| JD$_\odot$ | φ | Δm$_V$ | Δm$_B$ | Δm$_U$ |
|---|---|---|---|---|
| 2441510.3264 | 0.3131 | -0.312 | 0.610 | 0.107 |
| 2441510.3362 | 0.3140 | -0.318 | 0.642 | 0.116 |
| 2441510.3410 | 0.3144 | -0.354 | 0.625 | -0.058 |
| 2441510.3452 | 0.3148 | -0.327 | 0.599 | 0.026 |
| 2441510.3501 | 0.3153 | -0.341 | 0.616 | -0.161 |
| 2441510.3598 | 0.3161 | -0.357 | 0.632 | -0.126 |
| 2441510.3639 | 0.3165 | -0.313 | 0.554 | -0.114 |
| 2441510.3695 | 0.3170 | -0.335 | 0.638 | 0.071 |
| 2441537.2824 | 0.7362 | -0.288 | 0.583 | -0.126 |
| 2441537.2873 | 0.7367 | -0.390 | 0.595 | -0.104 |
| 2441537.2921 | 0.7371 | -0.384 | 0.533 | -0.009 |
| 2441537.2977 | 0.7376 | -0.417 | 0.554 | -0.034 |
| 2441537.3032 | 0.7381 | -0.382 | 0.567 | -0.096 |
| 2441537.3074 | 0.7385 | -0.321 | 0.503 | -0.057 |
| 2441858.3348 | 0.5956 | -0.257 | 0.719 | 0.024 |
| 2441858.3396 | 0.5960 | -0.175 | 0.849 | 0.120 |
| 2441858.3438 | 0.5964 | -0.120 | 0.757 | -0.012 |
| 2441858.3467 | 0.5967 | -0.169 | 0.800 | 0.000 |
| 2441858.3528 | 0.5972 | -0.152 | 0.793 | -0.062 |
| 2441858.3570 | 0.5976 | -0.146 | 0.788 | 0.042 |
| 2441858.3598 | 0.5979 | -0.200 | 0.832 | 0.025 |
| 2441858.3633 | 0.5982 | -0.210 | 0.873 | 0.074 |
| 2441858.3667 | 0.5985 | -0.210 | 0.891 | 0.030 |
| 2441858.3702 | 0.5988 | -0.229 | 0.767 | -0.030 |
| 2441888.3442 | 0.2932 | -0.347 | 0.614 | -0.044 |
| 2441888.3489 | 0.2936 | -0.349 | 0.625 | -0.161 |
| 2441888.3546 | 0.2941 | -0.377 | 0.663 | -0.099 |
| 2441888.3602 | 0.2946 | -0.343 | 0.599 | -0.022 |
| 2441888.3650 | 0.2950 | -0.379 | 0.605 | -0.044 |
| 2441890.3186 | 0.4706 | 0.124 | 1.080 | 0.420 |
| 2441890.3228 | 0.4710 | 0.120 | 1.100 | 0.445 |
| 2441890.3269 | 0.4714 | 0.119 | 1.111 | 0.394 |
| 2441890.3318 | 0.4718 | 0.122 | 1.066 | 0.389 |
| 2441890.3360 | 0.4722 | 0.111 | 1.047 | 0.421 |
| 2441890.3408 | 0.4726 | 0.120 | 1.086 | 0.423 |
| 2441890.3450 | 0.4730 | 0.131 | 1.070 | 0.418 |
| 2441890.3519 | 0.4736 | 0.103 | 1.125 | 0.501 |
| 2441890.3561 | 0.4740 | 0.147 | 1.096 | 0.401 |
| 2441890.3610 | 0.4744 | 0.137 | 1.091 | 0.414 |
| 2442221.3277 | 0.2250 | -0.326 | 0.679 | 0.002 |
| 2442221.3371 | 0.2259 | -0.287 | 0.693 | -0.064 |
| 2442221.3409 | 0.2262 | -0.272 | 0.709 | -0.070 |
| 2442221.3456 | 0.2266 | -0.292 | 0.691 | -0.005 |
| 2442221.3589 | 0.2278 | -0.331 | 0.658 | -0.112 |
| 2442221.3632 | 0.2282 | -0.317 | 0.659 | -0.095 |
| 2442221.3694 | 0.2288 | -0.268 | 0.654 | -0.038 |
| 2442221.3750 | 0.2293 | -0.311 | 0.642 | -0.065 |
| 2442221.3805 | 0.2298 | -0.335 | 0.614 | -0.043 |
| 2442221.3888 | 0.2305 | -0.313 | 0.694 | -0.110 |
| 2442221.3951 | 0.2311 | -0.305 | 0.671 | -0.007 |
| 2442221.3400 | 0.2261 | -0.295 | 0.667 | -0.117 |
| 2442221.4041 | 0.2319 | -0.283 | 0.696 | -0.037 |
| 2442221.4090 | 0.2323 | -0.311 | 0.704 | -0.021 |
| 2442244.2988 | 0.2899 | -0.297 | 0.646 | -0.072 |
| 2442244.3029 | 0.2903 | -0.329 | 0.627 | -0.072 |
| 2442244.3071 | 0.2906 | -0.290 | 0.599 | -0.093 |
| 2442244.3120 | 0.2911 | -0.280 | 0.611 | -0.064 |
| 2442244.3161 | 0.2915 | -0.319 | 0.630 | -0.063 |
| 2442244.3203 | 0.2918 | -0.340 | 0.621 | -0.034 |
| 2442244.3252 | 0.2923 | -0.331 | 0.612 | -0.054 |
| 2442244.3293 | 0.2926 | -0.341 | 0.614 | -0.088 |
| 2442244.3335 | 0.2930 | -0.349 | 0.641 | -0.038 |
| 2442244.3377 | 0.2934 | -0.348 | 0.627 | -0.004 |
| 2442244.3418 | 0.2938 | -0.318 | 0.656 | -0.024 |
| 2442248.3322 | 0.6525 | -0.303 | 0.713 | 0.002 |
| 2442248.3378 | 0.6530 | -0.331 | 0.673 | -0.004 |
| 2442250.2926 | 0.8287 | -0.259 | 0.685 | 0.001 |
| 2442250.2982 | 0.8292 | -0.277 | 0.688 | -0.018 |
| 2442250.3030 | 0.8296 | -0.313 | 0.692 | -0.026 |
| 2442250.3079 | 0.8301 | -0.289 | 0.645 | 0.004 |
| 2442250.3141 | 0.8306 | -0.259 | 0.644 | -0.100 |
| 2442250.3197 | 0.8311 | -0.304 | 0.656 | 0.016 |
| 2442250.3253 | 0.8316 | -0.308 | 0.674 | -0.045 |
| 2442250.3301 | 0.8321 | -0.323 | 0.687 | -0.073 |
| 2442250.3343 | 0.8324 | -0.287 | 0.699 | -0.010 |
| 2442250.3391 | 0.8329 | -0.286 | 0.703 | -0.014 |
| 2442250.3440 | 0.8333 | -0.255 | 0.710 | -0.023 |
| 2442250.3520 | 0.8340 | -0.260 | 0.690 | -0.040 |
| 2442250.3565 | 0.8344 | -0.326 | 0.679 | -0.032 |
| 2442273.2529 | 0.8926 | -0.098 | 0.870 | 0.328 |
| 2442273.2577 | 0.8930 | -0.119 | 0.826 | 0.133 |
| 2442273.2653 | 0.8937 | -0.106 | 0.874 | 0.188 |
| 2442273.2709 | 0.8942 | -0.069 | 0.912 | 0.292 |
| 2442273.2772 | 0.8948 | -0.033 | 0.889 | 0.185 |
| 2442273.2820 | 0.8952 | -0.139 | 0.784 | 0.377 |
| 2442273.2876 | 0.8957 | -0.117 | 0.892 | 0.198 |
| 2442273.2931 | 0.8962 | -0.059 | 0.805 | 0.150 |
| 2442273.2987 | 0.8967 | -0.140 | 0.813 | 0.215 |
| 2442273.3049 | 0.8973 | -0.055 | 0.854 | 0.175 |
| 2442653.2331 | 0.0490 | 0.119 | 1.183 | 0.576 |
| 2442653.2373 | 0.0494 | 0.118 | 1.203 | 0.534 |
| 2442653.2421 | 0.0498 | 0.006 | 0.965 | 0.273 |
| 2442653.2463 | 0.0502 | 0.121 | 1.013 | 0.290 |
| 2442663.2728 | 0.9515 | 0.279 | 1.300 | 0.608 |
| 2442663.2770 | 0.9518 | 0.307 | 1.267 | 0.613 |
| 2442663.2797 | 0.9521 | 0.138 | 1.108 | 0.523 |
| 2442663.2839 | 0.9525 | 0.253 | 1.288 | 0.582 |
| 2442663.2874 | 0.9528 | 0.296 | 1.244 | 0.579 |
| 2442663.2909 | 0.9531 | 0.243 | 1.228 | 0.518 |
| 2442663.2950 | 0.9535 | 0.172 | 1.134 | 0.469 |
| 2442663.2985 | 0.9538 | 0.285 | 1.155 | 0.441 |
| 2442663.3013 | 0.9540 | 0.235 | 1.270 | 0.421 |
| 2442663.3061 | 0.9545 | 0.295 | 1.117 | 0.410 |
| 2442930.4054 | 0.9640 | 0.216 | 1.191 | 0.541 |
| 2442930.4103 | 0.9645 | 0.296 | 1.268 | 0.578 |
| 2442930.4138 | 0.9648 | 0.198 | 1.284 | 0.569 |
| 2442930.4283 | 0.9661 | 0.291 | 1.218 | 0.640 |
| 2442930.4311 | 0.9663 | 0.278 | 1.318 | 0.621 |
| 2442930.4346 | 0.9666 | 0.225 | 1.241 | 0.635 |
| 2442930.4381 | 0.9670 | 0.320 | 1.284 | 0.591 |
| 2442930.4415 | 0.9673 | 0.248 | 1.243 | 0.598 |
| 2442930.4450 | 0.9676 | 0.229 | 1.197 | 0.572 |
| 2442930.4485 | 0.9679 | 0.262 | 1.222 | 0.562 |
| 2442930.4513 | 0.9681 | 0.203 | 1.197 | 0.566 |
| 2442930.4554 | 0.9685 | 0.236 | 1.203 | 0.593 |
| 2442930.4582 | 0.9688 | 0.305 | 1.276 | 0.693 |
| 2442930.4651 | 0.9694 | 0.346 | 1.249 | 0.698 |
| 2442930.4679 | 0.9696 | 0.300 | 1.373 | 0.680 |
| 2442933.4025 | 0.2334 | -0.216 | 0.753 | 0.096 |
| 2442933.4061 | 0.2337 | -0.263 | 0.656 | 0.111 |
| 2442933.4088 | 0.2340 | -0.226 | 0.712 | 0.030 |
| 2442933.4123 | 0.2343 | -0.230 | 0.725 | 0.068 |
| 2442933.4157 | 0.2346 | -0.277 | 0.713 | 0.037 |
| 2442933.4188 | 0.2349 | -0.290 | 0.682 | 0.067 |
| 2442933.4220 | 0.2352 | -0.265 | 0.657 | 0.024 |
| 2442955.3575 | 0.2070 | -0.168 | 0.796 | 0.122 |
| 2442955.3610 | 0.2073 | -0.169 | | 0.080 |
| 2442955.3651 | 0.2076 | -0.190 | 0.776 | 0.111 |
| 2442955.3686 | 0.2080 | -0.253 | 0.696 | 0.119 |
| 2442955.3721 | 0.2083 | -0.255 | 0.696 | 0.152 |
| 2442955.3762 | 0.2086 | -0.218 | 0.705 | 0.083 |
| 2442955.3797 | 0.2090 | -0.197 | 0.788 | 0.122 |
| 2442955.3832 | 0.2093 | -0.178 | 0.693 | 0.074 |
| 2442955.3860 | 0.2095 | -0.188 | 0.715 | 0.076 |
| 2442955.3894 | 0.2098 | -0.283 | 0.693 | 0.084 |
| 2442960.3485 | 0.6556 | -0.180 | 0.811 | 0.107 |
| 2442960.3526 | 0.6560 | -0.188 | 0.705 | 0.118 |
| 2442960.3561 | 0.6563 | -0.197 | 0.690 | 0.031 |
| 2442960.3596 | 0.6566 | -0.257 | 0.713 | 0.045 |
| 2442960.3631 | 0.6569 | -0.207 | 0.737 | 0.128 |
| 2442960.3665 | 0.6572 | -0.232 | 0.702 | 0.110 |
| 2442960.3700 | 0.6575 | -0.243 | 0.746 | 0.060 |
| 2442960.3735 | 0.6578 | -0.205 | 0.773 | 0.101 |
| 2442960.3769 | 0.6582 | -0.240 | 0.721 | 0.095 |
| 2442960.3804 | 0.6585 | -0.231 | 0.724 | 0.111 |
| 2442982.2767 | 0.6267 | -0.117 | 0.818 | 0.179 |
| 2442982.2795 | 0.6270 | -0.123 | 0.763 | 0.186 |
| 2442982.2830 | 0.6273 | -0.155 | 0.833 | 0.177 |
| 2442982.2865 | 0.6276 | -0.219 | 0.746 | 0.180 |
| 2442982.2892 | 0.6278 | -0.208 | 0.835 | 0.183 |
| 2442982.2927 | 0.6282 | -0.133 | 0.795 | 0.195 |
| 2442982.2955 | 0.6284 | -0.245 | 0.729 | 0.134 |
| 2442982.2990 | 0.6287 | -0.206 | 0.764 | 0.153 |
| 2442982.3059 | 0.6293 | -0.199 | 0.774 | 0.138 |
| 2442982.3094 | 0.6297 | -0.186 | 0.756 | 0.150 |
| 2442982.3122 | 0.6299 | -0.174 | 0.744 | 0.171 |
| 2442984.2741 | 0.8063 | -0.320 | 0.662 | -0.050 |
| 2442984.2768 | 0.8065 | -0.310 | 0.635 | -0.001 |
| 2442984.2803 | 0.8068 | -0.273 | 0.635 | -0.018 |
| 2442984.2845 | 0.8072 | -0.305 | 0.663 | 0.015 |
| 2442984.2935 | 0.8080 | -0.312 | 0.639 | -0.039 |
| 2442984.2970 | 0.8083 | -0.285 | 0.630 | -0.037 |
| 2442986.3014 | 0.9885 | 0.255 | 1.248 | 0.656 |
| 2442986.3138 | 0.9896 | 0.268 | 1.256 | 0.627 |
| 2442986.3172 | 0.9899 | 0.263 | 1.236 | 0.659 |
| 2442986.3207 | 0.9902 | 0.263 | 1.252 | 0.671 |
| 2442986.3242 | 0.9906 | 0.247 | 1.246 | 0.658 |
| 2442986.3311 | 0.9912 | 0.267 | 1.251 | 0.630 |
| 2442986.3381 | 0.9918 | 0.272 | 1.245 | 0.632 |
| 2442986.3408 | 0.9920 | 0.281 | 1.247 | 0.640 |
| 2442986.3443 | 0.9924 | 0.255 | 1.228 | 0.603 |
| 2442986.3478 | 0.9927 | 0.269 | 1.257 | 0.633 |
| 2442986.3513 | 0.9930 | 0.257 | 1.229 | 0.642 |
| 2442986.3540 | 0.9932 | 0.255 | 1.242 | 0.665 |
| 2442986.3575 | 0.9935 | 0.283 | 1.283 | 0.673 |
| 2442986.3610 | 0.9939 | 0.283 | 1.217 | 0.557 |
| 2442990.3326 | 0.3509 | -0.247 | 0.726 | 0.063 |
| 2442990.3361 | 0.3512 | -0.219 | 0.777 | 0.053 |
| 2442990.3395 | 0.3515 | -0.242 | 0.715 | 0.015 |
| 2443332.2930 | 0.0897 | -0.040 | 1.019 | 0.316 |
| 2443332.2965 | 0.0900 | -0.004 | 0.943 | 0.277 |
| 2443332.2986 | 0.0902 | -0.032 | 0.945 | 0.305 |
| 2443332.3055 | 0.0908 | -0.035 | 0.925 | 0.311 |
| 2443332.3083 | 0.0911 | 0.032 | 1.013 | 0.235 |
| 2443332.3118 | 0.0914 | -0.038 | 0.926 | 0.307 |
| 2443332.3152 | 0.0917 | -0.024 | 0.903 | 0.316 |
| 2443332.3194 | 0.0921 | -0.018 | 0.901 | 0.318 |
| 2443341.3147 | 0.9006 | 0.012 | 0.919 | 0.354 |
| 2443341.3189 | 0.9010 | -0.062 | 0.930 | 0.363 |
| 2443341.3224 | 0.9013 | -0.070 | 0.951 | 0.325 |
| 2443341.3265 | 0.9017 | -0.053 | 0.946 | 0.337 |
| 2443341.3307 | 0.9021 | -0.073 | 0.990 | 0.280 |
| 2443341.3335 | 0.9023 | -0.082 | 0.869 | 0.316 |
| 2443341.3370 | 0.9026 | -0.034 | 0.883 | 0.304 |
| 2443341.3418 | 0.9031 | -0.058 | 0.910 | 0.326 |
| 2443341.3453 | 0.9034 | -0.052 | 0.947 | 0.281 |
| 2443341.3488 | 0.9037 | -0.044 | 0.929 | 0.276 |
| 2443341.3522 | 0.9040 | -0.047 | 0.903 | 0.306 |
| 2443342.2897 | 0.9883 | 0.263 | 1.254 | 0.617 |
| 2443342.2932 | 0.9886 | 0.271 | 1.281 | 0.668 |
| 2443342.2967 | 0.9889 | 0.249 | 1.245 | 0.734 |
| 2443342.3002 | 0.9892 | 0.247 | 1.248 | 0.746 |
| 2443342.3036 | 0.9895 | 0.284 | 1.284 | 0.620 |
| 2443342.3071 | 0.9899 | 0.272 | 1.253 | 0.651 |
| 2443342.3006 | 0.9893 | 0.257 | 1.252 | 0.656 |
| 2443342.3040 | 0.9896 | 0.251 | 1.265 | 0.741 |
| 2443342.3175 | 0.9908 | 0.247 | 1.259 | 0.615 |
| 2443342.3203 | 0.9910 | 0.282 | 1.239 | 0.655 |
| 2443346.2940 | 0.3482 | -0.284 | 0.651 | 0.011 |
| 2443346.2975 | 0.3485 | -0.291 | 0.697 | 0.009 |
| 2443346.3010 | 0.3489 | -0.270 | 0.666 | 0.052 |
| 2443346.3072 | 0.3494 | -0.286 | 0.664 | 0.037 |
| 2443346.3107 | 0.3497 | -0.278 | 0.679 | 0.020 |
| 2443346.3141 | 0.3500 | -0.295 | 0.670 | -0.017 |
| 2443346.3176 | 0.3504 | -0.302 | 0.672 | 0.046 |
| 2443346.3211 | 0.3507 | -0.284 | 0.712 | 0.026 |
| 2443346.3246 | 0.3510 | -0.257 | 0.716 | 0.092 |
| 2443346.3280 | 0.3513 | -0.279 | 0.750 | 0.108 |
| 2443391.2517 | 0.3895 | -0.076 | 0.876 | 0.070 |
| 2443391.2545 | 0.3897 | -0.093 | 0.854 | 0.153 |
| 2443391.2579 | 0.3900 | -0.133 | 0.889 | 0.113 |
| 2443391.2614 | 0.3904 | -0.137 | 0.768 | 0.133 |
| 2443391.2642 | 0.3906 | -0.152 | 0.834 | 0.168 |
| 2443391.2676 | 0.3909 | -0.204 | 0.845 | 0.061 |
| 2443391.2704 | 0.3912 | -0.175 | 0.904 | 0.050 |
| 2443420.1889 | 0.9906 | 0.285 | 1.253 | 0.633 |
| 2443420.1924 | 0.9910 | 0.296 | 1.271 | 0.678 |
| 2443420.1979 | 0.9915 | 0.305 | 1.264 | 0.692 |
| 2443420.2021 | 0.9918 | 0.287 | 1.258 | 0.662 |
| 2443420.2055 | 0.9921 | 0.286 | 1.251 | 0.641 |
| 2443420.2083 | 0.9924 | 0.286 | 1.258 | 0.652 |
| 2443420.2125 | 0.9928 | 0.287 | 1.260 | 0.639 |
| 2443420.2167 | 0.9931 | 0.292 | 1.251 | 0.611 |
| 2443420.2194 | 0.9934 | 0.283 | 1.256 | 0.655 |
| 2443423.1873 | 0.2602 | -0.201 | 0.757 | 0.054 |
| 2443423.1908 | 0.2605 | -0.294 | 0.666 | 0.052 |
| 2443423.1942 | 0.2608 | -0.295 | 0.695 | 0.058 |
| 2443423.1970 | 0.2610 | -0.287 | 0.655 | 0.079 |
| 2443423.1998 | 0.2613 | -0.296 | 0.674 | 0.020 |
| 2443423.2033 | 0.2616 | -0.292 | 0.690 | 0.005 |
| 2443423.2067 | 0.2619 | -0.291 | 0.676 | 0.006 |
| 2443423.2102 | 0.2622 | -0.305 | 0.657 | 0.009 |
| 2443423.2130 | 0.2625 | -0.286 | 0.666 | -0.019 |
| 2443423.2158 | 0.2627 | -0.302 | 0.657 | -0.001 |
| 2443424.1928 | 0.3506 | -0.249 | 0.700 | 0.097 |
| 2443424.1963 | 0.3509 | -0.246 | 0.696 | 0.076 |
| 2443424.1984 | 0.3511 | -0.251 | 0.703 | 0.051 |
| 2443424.2026 | 0.3514 | -0.260 | 0.710 | 0.066 |
| 2443424.2060 | 0.3517 | -0.259 | 0.700 | 0.063 |
| 2443424.2088 | 0.3520 | -0.249 | 0.692 | 0.027 |
| 2443424.2109 | 0.3522 | -0.259 | 0.695 | 0.023 |
| 2443424.2144 | 0.3525 | -0.268 | 0.691 | 0.089 |
| 2443424.2185 | 0.3529 | -0.267 | 0.677 | 0.027 |
| 2443424.2213 | 0.3531 | -0.245 | 0.673 | 0.025 |
| 2443425.1941 | 0.4406 | -0.002 | 0.977 | 0.342 |
| 2443425.1976 | 0.4409 | 0.011 | 0.972 | 0.264 |
| 2443425.2004 | 0.4411 | 0.001 | 0.955 | 0.288 |
| 2443425.2046 | 0.4415 | 0.014 | 0.948 | 0.287 |
| 2443425.2073 | 0.4417 | 0.006 | 0.966 | 0.314 |
| 2443425.2108 | 0.4421 | 0.011 | 0.971 | 0.323 |
| 2443425.2136 | 0.4423 | -0.009 | 0.930 | 0.303 |
| 2443425.2164 | 0.4426 | 0.006 | 0.947 | 0.293 |
| 2443425.2191 | 0.4428 | -0.011 | 0.966 | 0.293 |
| 2443425.2205 | 0.4429 | 0.009 | 0.961 | 0.278 |
| 2443429.1820 | 0.7990 | -0.320 | 0.633 | -0.030 |
| 2443429.1848 | 0.7993 | -0.313 | 0.619 | -0.009 |
| 2443429.1890 | 0.7997 | -0.333 | 0.635 | -0.010 |
| 2443429.1918 | 0.7999 | -0.346 | 0.637 | -0.059 |
| 2443429.1952 | 0.8002 | -0.330 | 0.638 | -0.051 |
| 2443429.1980 | 0.8005 | -0.333 | 0.624 | -0.058 |
| 2443429.2006 | 0.8007 | -0.333 | 0.623 | -0.063 |
| 2443429.2043 | 0.8010 | -0.333 | 0.615 | -0.087 |
| 2443429.2070 | 0.8013 | -0.332 | 0.607 | -0.059 |
| 2443429.2098 | 0.8015 | -0.335 | 0.616 | -0.081 |
| 2443430.1798 | 0.8887 | -0.125 | 0.840 | 0.234 |
| 2443430.1826 | 0.8890 | -0.110 | 0.868 | 0.240 |
| 2443430.1854 | 0.8892 | -0.111 | 0.866 | 0.178 |
| 2443430.1882 | 0.8895 | -0.109 | 0.851 | 0.202 |
| 2443430.1910 | 0.8897 | -0.107 | 0.885 | 0.189 |
| 2443430.1937 | 0.8900 | -0.129 | 0.865 | 0.185 |
| 2443430.1965 | 0.8902 | -0.094 | 0.860 | 0.195 |
| 2443430.1993 | 0.8905 | -0.082 | 0.861 | 0.185 |
| 2443430.2021 | 0.8907 | -0.070 | 0.875 | 0.201 |
| 2443430.2048 | 0.8910 | -0.094 | 0.844 | 0.194 |
| 2443666.3766 | 0.1205 | -0.146 | 0.759 | -0.020 |
| 2443666.3801 | 0.1208 | -0.119 | 0.839 | 0.184 |
| 2443666.4010 | 0.1226 | -0.110 | | 0.053 |
| 2443666.4044 | 0.1230 | -0.117 | 0.855 | |
| 2443666.4079 | 0.1233 | -0.156 | 0.840 | 0.211 |
| 2443666.4114 | 0.1236 | -0.133 | 0.876 | 0.108 |
| 2443666.4148 | 0.1239 | -0.150 | 0.887 | 0.158 |
| 2443666.4183 | 0.1242 | -0.158 | 0.855 | 0.143 |
| 2443666.4218 | 0.1245 | -0.115 | 0.841 | 0.116 |
| 2443666.4253 | 0.1248 | -0.135 | 0.852 | 0.205 |
| 2443666.4287 | 0.1251 | -0.144 | 0.838 | 0.296 |
| 2443666.4315 | 0.1254 | -0.150 | 0.839 | 0.219 |
| 2443666.4350 | 0.1257 | -0.104 | 0.817 | 0.212 |
| 2443666.4385 | 0.1260 | -0.124 | 0.824 | 0.142 |
| 2443666.4433 | 0.1264 | -0.135 | 0.768 | 0.191 |
| 2443666.4468 | 0.1268 | -0.176 | 0.848 | 0.189 |
| 2443666.4496 | 0.1270 | -0.138 | 0.854 | 0.229 |
| 2443666.4530 | 0.1273 | -0.144 | 0.785 | 0.203 |
| 2443670.3395 | 0.4767 | 0.122 | 1.117 | 0.413 |
| 2443670.3423 | 0.4769 | 0.115 | 1.097 | 0.424 |
| 2443670.3458 | 0.4772 | 0.132 | 1.079 | 0.472 |
| 2443670.3500 | 0.4776 | 0.126 | 1.121 | 0.474 |
| 2443670.3527 | 0.4779 | 0.152 | 1.115 | 0.460 |
| 2443670.3569 | 0.4782 | 0.173 | 1.105 | 0.479 |
| 2443670.3597 | 0.4785 | 0.145 | 1.122 | 0.442 |
| 2443670.3638 | 0.4789 | 0.155 | 1.118 | 0.461 |
| 2443670.3673 | 0.4792 | 0.144 | 1.129 | 0.487 |
| 2443670.3763 | 0.4800 | 0.147 | 1.120 | 0.442 |
| 2443670.3805 | 0.4804 | 0.152 | 1.137 | 0.455 |
| 2443670.3840 | 0.4807 | 0.152 | 1.101 | 0.434 |
| 2443670.3868 | 0.4809 | 0.139 | 1.122 | 0.502 |
| 2443670.3895 | 0.4812 | 0.154 | 1.098 | 0.504 |
| 2443670.3930 | 0.4815 | 0.131 | 1.070 | 0.470 |
| 2443670.3958 | 0.4817 | 0.146 | 1.100 | 0.470 |
| 2443670.3993 | 0.4821 | 0.128 | 1.116 | 0.472 |
| 2443670.4020 | 0.4823 | 0.160 | 1.104 | 0.493 |
| 2443670.4083 | 0.4829 | 0.154 | 1.132 | 0.443 |
| 2443670.4111 | 0.4831 | 0.148 | 1.121 | 0.456 |
| 2443671.3759 | 0.5698 | -0.106 | 0.877 | 0.212 |
| 2443671.3793 | 0.5701 | -0.118 | 0.871 | 0.198 |
| 2443671.3828 | 0.5705 | -0.116 | 0.883 | 0.201 |
| 2443671.3863 | 0.5708 | -0.120 | 0.886 | 0.196 |
| 2443671.3890 | 0.5710 | -0.109 | 0.879 | 0.193 |
| 2443671.3925 | 0.5713 | -0.114 | 0.907 | 0.175 |
| 2443671.3988 | 0.5719 | -0.107 | 0.907 | 0.175 |
| 2443671.4015 | 0.5721 | -0.113 | 0.890 | 0.192 |
| 2443671.4050 | 0.5725 | -0.116 | 0.880 | 0.188 |
| 2443671.4085 | 0.5728 | -0.117 | 0.860 | 0.200 |
| 2443671.4120 | 0.5731 | -0.108 | 0.876 | 0.196 |
| 2443671.4147 | 0.5733 | -0.104 | 0.880 | 0.196 |
| 2443671.4182 | 0.5736 | -0.131 | 0.856 | 0.200 |
| 2443671.4224 | 0.5740 | -0.127 | 0.872 | 0.209 |
| 2443671.4252 | 0.5743 | -0.122 | 0.877 | 0.148 |
| 2443671.4321 | 0.5749 | -0.123 | 0.887 | 0.167 |
| 2443671.4349 | 0.5751 | -0.107 | 0.877 | 0.143 |
| 2443671.4377 | 0.5754 | -0.107 | 0.875 | 0.183 |
| 2443671.4404 | 0.5756 | -0.116 | 0.852 | 0.180 |
| 2443671.4432 | 0.5759 | -0.121 | 0.875 | 0.161 |
| 2443672.4161 | 0.6633 | -0.252 | 0.744 | 0.121 |
| 2443672.4245 | 0.6641 | -0.239 | 0.739 | 0.120 |
| 2443672.4265 | 0.6643 | -0.245 | 0.719 | 0.087 |
| 2443672.4293 | 0.6645 | -0.245 | 0.720 | 0.115 |
| 2443672.4314 | 0.6647 | -0.253 | 0.722 | 0.109 |
| 2443672.4342 | 0.6650 | -0.232 | 0.715 | 0.078 |
| 2443672.4370 | 0.6652 | -0.242 | 0.734 | 0.100 |
| 2443673.4424 | 0.7556 | -0.323 | 0.655 | 0.033 |
| 2443673.4452 | 0.7558 | -0.314 | 0.654 | 0.058 |
| 2443673.4480 | 0.7561 | -0.322 | 0.651 | -0.017 |
| 2443673.4508 | 0.7564 | -0.307 | 0.674 | -0.029 |
| 2443673.4535 | 0.7566 | -0.306 | 0.646 | 0.016 |
| 2443673.4563 | 0.7568 | -0.312 | 0.636 | 0.033 |
| 2443673.4591 | 0.7571 | -0.311 | 0.643 | 0.009 |
| 2443673.4619 | 0.7574 | -0.310 | 0.641 | 0.012 |
| 2443673.4646 | 0.7576 | -0.316 | 0.649 | 0.012 |
| 2443673.4674 | 0.7578 | -0.322 | 0.659 | -0.005 |
| 2443687.3526 | 0.0060 | 0.276 | 1.265 | 0.663 |
| 2443687.3568 | 0.0064 | 0.264 | 1.257 | 0.622 |
| 2443687.3596 | 0.0066 | 0.265 | 1.227 | 0.643 |
| 2443687.3624 | 0.0069 | 0.291 | 1.264 | 0.643 |
| 2443687.3651 | 0.0071 | 0.267 | 1.278 | 0.636 |
| 2443687.3679 | 0.0074 | 0.262 | 1.261 | 0.672 |
| 2443687.3707 | 0.0076 | 0.272 | 1.264 | 0.647 |
| 2443687.3735 | 0.0079 | 0.279 | 1.254 | 0.650 |
| 2443687.3762 | 0.0081 | 0.283 | 1.270 | 0.659 |
| 2443688.3540 | 0.0960 | -0.039 | 0.921 | 0.286 |
| 2443688.3631 | 0.0968 | -0.050 | 0.905 | 0.293 |
| 2443688.3665 | 0.0971 | -0.042 | 0.911 | 0.268 |
| 2443688.3700 | 0.0974 | -0.044 | 0.900 | 0.278 |
| 2443688.3728 | 0.0977 | -0.056 | 0.911 | 0.314 |
| 2443688.3762 | 0.0980 | -0.054 | 0.920 | 0.279 |
| 2443688.3790 | 0.0982 | -0.037 | 0.937 | 0.294 |
| 2443688.3818 | 0.0985 | -0.036 | 0.938 | 0.296 |
| 2443688.3853 | 0.0988 | -0.041 | 0.917 | 0.296 |
| 2443688.3881 | 0.0991 | -0.052 | 0.899 | 0.247 |
| 2443688.3908 | 0.0993 | -0.041 | 0.904 | 0.254 |
| 2443695.3826 | 0.7278 | -0.304 | 0.664 | -0.001 |
| 2443695.3861 | 0.7281 | -0.317 | 0.648 | -0.004 |
| 2443695.3898 | 0.7284 | -0.307 | 0.641 | -0.004 |
| 2443695.3923 | 0.7287 | -0.308 | 0.638 | -0.011 |
| 2443695.3951 | 0.7289 | -0.308 | 0.653 | 0.001 |
| 2443695.3979 | 0.7292 | -0.291 | 0.646 | 0.000 |
| 2443695.4007 | 0.7294 | -0.312 | 0.673 | -0.012 |
| 2443696.3972 | 0.8190 | -0.312 | 0.670 | 0.051 |
| 2443696.4000 | 0.8193 | -0.310 | 0.668 | 0.042 |
| 2443696.4034 | 0.8196 | -0.302 | 0.661 | 0.012 |
| 2443696.4062 | 0.8198 | -0.290 | 0.676 | 0.041 |
| 2443696.4083 | 0.8200 | -0.312 | 0.682 | 0.018 |
| 2443696.4111 | 0.8203 | -0.296 | 0.707 | 0.007 |
| 2443696.4138 | 0.8205 | -0.311 | 0.682 | 0.026 |
| 2443697.3972 | 0.9089 | -0.020 | 0.943 | 0.311 |
| 2443697.4000 | 0.9091 | 0.088 | 1.048 | 0.423 |
| 2443697.4027 | 0.9094 | 0.010 | 0.938 | 0.325 |
| 2443697.4055 | 0.9096 | -0.004 | 0.957 | 0.336 |
| 2443697.4083 | 0.9099 | -0.007 | 0.974 | 0.310 |
| 2443697.4118 | 0.9102 | -0.022 | 0.972 | 0.315 |
| 2443697.4145 | 0.9104 | -0.022 | 0.975 | 0.334 |
| 2443697.4222 | 0.9111 | -0.017 | 0.978 | 0.300 |
| 2443697.4305 | 0.9119 | 0.009 | 1.010 | 0.399 |
| 2443698.4180 | 0.0007 | 0.272 | 1.271 | 0.652 |
| 2443698.4198 | 0.0008 | 0.279 | 1.270 | 0.642 |
| 2443698.4229 | 0.0011 | 0.271 | 1.250 | 0.625 |
| 2443698.4257 | 0.0013 | 0.278 | 1.259 | 0.692 |
| 2443698.4284 | 0.0016 | 0.259 | 1.222 | 0.571 |
| 2443698.4312 | 0.0018 | 0.259 | 1.222 | 0.547 |
| 2443698.4326 | 0.0020 | 0.256 | 1.230 | 0.611 |
| 2443720.2847 | 0.9663 | 0.203 | 1.191 | 0.570 |
| 2443720.2875 | 0.9665 | 0.216 | 1.201 | 0.522 |
| 2443720.2902 | 0.9667 | 0.228 | 1.171 | 0.436 |
| 2443720.2937 | 0.9671 | 0.208 | 1.165 | 0.514 |
| 2443720.2972 | 0.9674 | 0.204 | 1.198 | 0.553 |
| 2443720.3000 | 0.9676 | 0.187 | 1.179 | 0.642 |
| 2443720.3111 | 0.9686 | 0.224 | 1.209 | 0.509 |
| 2443720.3166 | 0.9691 | 0.224 | 1.167 | 0.509 |
| 2443720.3194 | 0.9694 | 0.198 | 1.177 | 0.679 |
| 2443720.3222 | 0.9696 | 0.201 | 1.180 | 0.552 |
| 2443720.3250 | 0.9699 | 0.209 | 1.185 | 0.575 |
| 2443720.3277 | 0.9701 | 0.328 | 1.218 | 0.503 |
| 2443720.3305 | 0.9704 | 0.228 | 1.218 | 0.502 |
| 2443720.3340 | 0.9707 | 0.211 | 1.204 | 0.494 |
| 2443727.2927 | 0.5962 | -0.144 | 0.824 | 0.220 |
| 2443727.2955 | 0.5965 | -0.155 | 0.814 | 0.206 |
| 2443727.2983 | 0.5967 | -0.060 | 0.869 | 0.200 |
| 2443727.3018 | 0.5970 | -0.150 | 0.822 | 0.133 |
| 2443727.3045 | 0.5973 | -0.123 | 0.815 | 0.099 |
| 2443727.3080 | 0.5976 | -0.136 | 0.806 | 0.012 |
| 2443727.3108 | 0.5978 | -0.152 | 0.886 | 0.066 |
| 2443727.3136 | 0.5981 | -0.170 | 0.800 | 0.081 |
| 2443727.3163 | 0.5983 | -0.143 | 0.809 | 0.117 |
| 2443727.3171 | 0.5984 | -0.143 | 0.808 | -0.006 |
| 2443730.3158 | 0.8679 | -0.191 | 0.742 | 0.074 |
| 2443730.3186 | 0.8682 | -0.199 | 0.747 | 0.105 |
| 2443730.3214 | 0.8685 | -0.139 | 0.740 | 0.113 |
| 2443730.3256 | 0.8688 | -0.179 | 0.738 | 0.089 |
| 2443730.3276 | 0.8690 | -0.194 | 0.756 | 0.094 |
| 2443730.3304 | 0.8693 | -0.209 | 0.755 | 0.086 |
| 2443730.3332 | 0.8695 | -0.219 | 0.734 | 0.070 |
| 2443730.3360 | 0.8698 | -0.209 | 0.744 | 0.082 |
| 2443730.3388 | 0.8700 | -0.126 | 0.806 | 0.147 |
| 2443730.3415 | 0.8703 | -0.216 | 0.740 | 0.066 |
| 2443731.3118 | 0.9575 | 0.170 | 1.133 | 0.539 |
| 2443731.3146 | 0.9577 | 0.170 | 1.141 | 0.532 |
| 2443731.3173 | 0.9580 | 0.162 | 1.147 | 0.481 |
| 2443731.3201 | 0.9582 | 0.174 | 1.179 | 0.482 |
| 2443731.3229 | 0.9585 | 0.137 | 1.157 | 0.422 |
| 2443731.3264 | 0.9588 | 0.148 | 1.163 | 0.425 |
| 2443731.3291 | 0.9590 | 0.181 | 1.170 | 0.362 |
| 2443731.3326 | 0.9593 | | 1.132 | 0.448 |
| 2443731.3354 | 0.9596 | 0.182 | 1.172 | 0.547 |
| 2443731.3382 | 0.9599 | 0.185 | 1.172 | 0.514 |
| 2443731.3416 | 0.9602 | 0.147 | 1.144 | 0.428 |
| 2443731.3444 | 0.9604 | 0.168 | 1.126 | 0.330 |
| 2443751.2404 | 0.7489 | -0.288 | 0.645 | 0.077 |
| 2443751.2431 | 0.7491 | -0.301 | 0.640 | 0.099 |
| 2443751.2459 | 0.7494 | -0.275 | 0.641 | 0.101 |
| 2443751.2487 | 0.7496 | -0.300 | 0.647 | 0.045 |
| 2443751.2515 | 0.7499 | -0.294 | 0.694 | -0.055 |
| 2443751.2536 | 0.7500 | -0.299 | 0.697 | 0.006 |
| 2443751.2563 | 0.7503 | -0.265 | 0.675 | -0.054 |
| 2443751.2591 | 0.7505 | -0.273 | 0.624 | 0.071 |
| 2443751.2619 | 0.7508 | -0.283 | 0.608 | -0.059 |
| 2443751.2647 | 0.7510 | -0.277 | 0.650 | -0.014 |
| 2443779.2130 | 0.2633 | -0.255 | 0.636 | -0.001 |
| 2443779.2158 | 0.2636 | -0.322 | 0.641 | 0.005 |
| 2443779.2192 | 0.2639 | -0.235 | 0.725 | 0.028 |
| 2443779.2220 | 0.2641 | -0.310 | 0.642 | -0.029 |
| 2443779.2248 | 0.2644 | -0.319 | 0.648 | -0.016 |
| 2443779.2276 | 0.2646 | -0.318 | 0.622 | -0.003 |
| 2443779.2304 | 0.2649 | -0.323 | 0.632 | -0.009 |
| 2443781.2326 | 0.4449 | 0.010 | 0.991 | 0.302 |
| 2443781.2354 | 0.4451 | 0.031 | 0.968 | 0.320 |
| 2443781.2382 | 0.4454 | 0.044 | 0.961 | 0.310 |
| 2443781.2417 | 0.4457 | 0.034 | 0.985 | 0.335 |
| 2443781.2451 | 0.4460 | 0.020 | 0.973 | 0.321 |
| 2443787.2124 | 0.9824 | 0.247 | 1.202 | 0.511 |
| 2443787.2145 | 0.9826 | - | 1.239 | 0.543 |
| 2443787.2173 | 0.9828 | 0.239 | 1.172 | 0.527 |
| 2443787.2200 | 0.9831 | 0.236 | 1.210 | 0.537 |
| 2443787.2228 | 0.9833 | 0.244 | 1.199 | 0.527 |
| 2443787.2256 | 0.9836 | 0.241 | 1.203 | 0.540 |
| 2443787.2284 | 0.9838 | 0.227 | 1.213 | 0.502 |
| 2443787.2305 | 0.9840 | 0.232 | 1.189 | 0.439 |
| 2444049.3089 | 0.5422 | - | - | 0.371 |
| 2444049.3117 | 0.5425 | 0.047 | 0.975 | 0.385 |
| 2444049.3158 | 0.5429 | 0.020 | 1.008 | 0.360 |
| 2444049.3193 | 0.5432 | 0.031 | 0.996 | 0.394 |
| 2444049.3228 | 0.5435 | 0.038 | 1.011 | 0.422 |
| 2444049.3262 | 0.5438 | 0.003 | 0.989 | 0.335 |
| 2444055.3387 | 0.0843 | 0.005 | 0.969 | 0.277 |
| 2444055.3422 | 0.0846 | 0.010 | 0.979 | 0.336 |
| 2444055.3450 | 0.0848 | -0.007 | 0.953 | 0.335 |
| 2444055.3478 | 0.0851 | -0.014 | 0.904 | 0.249 |
| 2444086.3507 | 0.8719 | -0.205 | 0.741 | 0.076 |
| 2444086.3534 | 0.8722 | -0.196 | 0.768 | 0.137 |
| 2444086.3562 | 0.8724 | -0.182 | 0.756 | 0.087 |
| 2444086.3597 | 0.8727 | -0.220 | 0.743 | 0.118 |
| 2444434.3139 | 0.1503 | -0.142 | 0.838 | 0.142 |
| 2444434.3174 | 0.1506 | -0.174 | 0.829 | 0.168 |
| 2444434.3216 | 0.1510 | -0.145 | 0.827 | 0.124 |
| 2444434.3285 | 0.1516 | -0.181 | 0.804 | 0.126 |
| 2444434.3320 | 0.1520 | -0.132 | 0.793 | 0.125 |
| 2444434.3348 | 0.1522 | -0.143 | 0.857 | 0.163 |
| 2444434.3383 | 0.1525 | -0.145 | 0.843 | 0.124 |
| 2444434.3410 | 0.1528 | -0.150 | 0.823 | 0.187 |
| 2444434.3514 | 0.1537 | -0.155 | 0.836 | 0.131 |
| 2444434.3549 | 0.1540 | -0.137 | 0.840 | 0.090 |
| 2444437.3439 | 0.4227 | -0.023 | 0.915 | 0.238 |
| 2444437.3467 | 0.4229 | 0.001 | 0.952 | 0.269 |
| 2444437.3495 | 0.4232 | -0.009 | 0.960 | 0.287 |
| 2444437.3522 | 0.4234 | -0.035 | 0.941 | 0.271 |
| 2444437.3550 | 0.4237 | -0.060 | 0.889 | 0.198 |
| 2444437.3578 | 0.4239 | -0.057 | 0.955 | 0.259 |
| 2444437.3606 | 0.4242 | -0.009 | 0.924 | 0.188 |
| 2444437.3634 | 0.4244 | -0.016 | 0.964 | 0.197 |
| 2444437.3661 | 0.4247 | -0.034 | 0.975 | 0.249 |
| 2444437.3689 | 0.4249 | -0.023 | 0.979 | 0.244 |
| 2444438.3106 | 0.5096 | 0.190 | 1.209 | 0.534 |
| 2444438.3140 | 0.5099 | 0.131 | 1.196 | 0.520 |
| 2444438.3168 | 0.5101 | 0.224 | 1.227 | 0.509 |
| 2444438.3196 | 0.5104 | 0.218 | 1.235 | 0.531 |
| 2444438.3245 | 0.5108 | 0.250 | 1.225 | 0.428 |
| 2444438.3272 | 0.5111 | 0.227 | 1.112 | 0.427 |
| 2444438.3300 | 0.5113 | 0.285 | 1.181 | 0.383 |
| 2444438.3328 | 0.5116 | 0.234 | 1.101 | 0.474 |
| 2444438.3363 | 0.5119 | 0.272 | 1.209 | 0.457 |
| 2444438.3390 | 0.5121 | 0.218 | 1.180 | 0.519 |
| 2444442.3919 | 0.8765 | -0.138 | 0.876 | 0.128 |
| 2444442.3954 | 0.8768 | -0.122 | 0.849 | 0.127 |
| 2444442.3996 | 0.8771 | -0.159 | 0.848 | 0.092 |
| 2444442.4093 | 0.8780 | -0.095 | 0.846 | |
| 2444442.4128 | 0.8783 | -0.068 | 0.893 | |
| 2444459.2614 | 0.3929 | -0.124 | 0.726 | 0.199 |
| 2444459.2678 | 0.3934 | -0.126 | 0.807 | 0.204 |
| 2444459.2706 | 0.3937 | -0.202 | 0.804 | 0.158 |
| 2444459.2734 | 0.3939 | -0.110 | 0.879 | 0.224 |
| 2444459.2769 | 0.3942 | -0.110 | 0.844 | 0.128 |
| 2444459.2796 | 0.3945 | -0.209 | 0.722 | 0.101 |
| 2444459.2824 | 0.3947 | -0.210 | 0.728 | 0.140 |
| 2444494.2041 | 0.5339 | 0.056 | 1.029 | 0.362 |
| 2444494.2111 | 0.5345 | 0.036 | 0.994 | 0.394 |
| 2444494.2146 | 0.5348 | 0.024 | 1.002 | 0.291 |
| 2444494.2194 | 0.5352 | 0.036 | 0.972 | 0.354 |
| 2444494.2222 | 0.5355 | 0.028 | 1.000 | 0.362 |
| 2444494.2257 | 0.5358 | 0.035 | 1.026 | 0.366 |
| 2444494.2291 | 0.5361 | 0.012 | 0.988 | 0.324 |
| 2444494.2319 | 0.5364 | 0.015 | 0.990 | 0.316 |
| 2444494.2347 | 0.5366 | 0.007 | 0.967 | 0.410 |
| 2444494.2382 | 0.5369 | 0.019 | 0.951 | 0.295 |
| 2444494.2409 | 0.5372 | 0.039 | 0.957 | 0.271 |
| 2444494.2437 | 0.5374 | 0.051 | 0.941 | 0.289 |
| 2444754.3630 | 0.9195 | 0.005 | 1.020 | 0.301 |
| 2444754.3678 | 0.9200 | 0.093 | 0.963 | 0.272 |
| 2444754.3713 | 0.9203 | -0.025 | 0.982 | 0.293 |
| 2444754.3775 | 0.9208 | -0.008 | 0.995 | 0.317 |
| 2444754.3796 | 0.9210 | -0.013 | 0.985 | 0.335 |
| 2444754.3831 | 0.9213 | -0.003 | 0.972 | 0.352 |
| 2444754.3866 | 0.9217 | 0.019 | 0.981 | 0.336 |
| 2444754.3900 | 0.9220 | 0.006 | 0.959 | 0.363 |
| 2444754.3984 | 0.9227 | 0.015 | 0.981 | 0.381 |
| 2444754.4032 | 0.9231 | 0.037 | 0.979 | 0.367 |
| 2444754.4060 | 0.9234 | 0.051 | 0.994 | 0.393 |
| 2444754.4095 | 0.9237 | 0.012 | 1.016 | 0.421 |
| 2444754.4130 | 0.9240 | 0.012 | 0.993 | 0.412 |
| 2444754.4164 | 0.9243 | 0.034 | 1.008 | 0.382 |
| 2444754.4206 | 0.9247 | 0.025 | 1.006 | 0.413 |
| 2444754.4234 | 0.9250 | 0.027 | 1.006 | 0.387 |
| 2444754.4268 | 0.9253 | 0.055 | 1.016 | 0.358 |
| 2444754.4296 | 0.9255 | 0.059 | 1.026 | 0.384 |
| 2444754.4331 | 0.9258 | 0.017 | 1.010 | 0.414 |
| 2444754.4385 | 0.9263 | 0.018 | 1.042 | 0.425 |
| 2444754.4421 | 0.9266 | 0.034 | 0.995 | 0.386 |
| 2444754.4463 | 0.9270 | 0.040 | 1.030 | 0.400 |
| 2444754.4498 | 0.9273 | 0.053 | 1.070 | 0.436 |
| 2444754.4592 | 0.9282 | 0.050 | 0.998 | 0.394 |
| 2444754.4553 | 0.9278 | 0.067 | 1.008 | 0.396 |
| 2444754.4595 | 0.9282 | 0.055 | 1.025 | 0.379 |
| 2444754.4630 | 0.9285 | 0.061 | 1.017 | 0.399 |
| 2444754.4671 | 0.9289 | 0.053 | 1.008 | 0.399 |
| 2444754.4699 | 0.9291 | 0.077 | 1.040 | 0.446 |
| 2444754.4720 | 0.9293 | 0.060 | 1.050 | 0.417 |
| 2444765.4120 | 0.9127 | -0.059 | 0.899 | 0.259 |
| 2444765.4189 | 0.9133 | -0.045 | 0.926 | 0.263 |
| 2444765.4224 | 0.9137 | -0.039 | 0.965 | 0.230 |
| 2444765.4286 | 0.9142 | -0.043 | 0.920 | 0.242 |
| 2444765.4314 | 0.9145 | -0.036 | 0.956 | 0.244 |
| 2444765.4349 | 0.9148 | -0.033 | 0.961 | 0.261 |
| 2444765.4377 | 0.9150 | -0.018 | 0.962 | 0.267 |
| 2444765.4418 | 0.9154 | -0.036 | 0.940 | 0.263 |
| 2444765.4446 | 0.9157 | -0.036 | 0.927 | 0.273 |
| 2444765.4481 | 0.9160 | -0.043 | 0.934 | 0.258 |
| 2444765.4516 | 0.9163 | -0.032 | 0.957 | 0.270 |
| 2444765.4543 | 0.9165 | -0.043 | 0.954 | 0.273 |
| 2444765.4571 | 0.9168 | -0.026 | 0.910 | 0.258 |
| 2444765.4606 | 0.9171 | -0.033 | 1.031 | 0.280 |
| 2444765.4634 | 0.9173 | -0.056 | 0.969 | 0.297 |
| 2444765.4661 | 0.9176 | -0.004 | 0.980 | 0.247 |
| 2444765.4689 | 0.9178 | -0.026 | 0.923 | 0.286 |
| 2444765.4717 | 0.9181 | -0.020 | 0.926 | 0.289 |
| 2444809.2879 | 0.8567 | -0.201 | 0.731 | 0.085 |
| 2444809.2899 | 0.8569 | -0.211 | 0.740 | 0.113 |
| 2444809.2934 | 0.8572 | -0.238 | 0.741 | 0.110 |
| 2444809.2990 | 0.8577 | -0.226 | 0.761 | 0.108 |
| 2444809.3024 | 0.8580 | -0.219 | 0.752 | 0.085 |
| 2444809.3059 | 0.8583 | -0.212 | 0.771 | 0.113 |
| 2444809.3087 | 0.8586 | -0.194 | 0.764 | 0.154 |
| 2444809.3115 | 0.8589 | -0.199 | 0.744 | 0.076 |
| 2444809.3156 | 0.8592 | -0.240 | 0.729 | 0.072 |
| 2444809.3184 | 0.8595 | -0.211 | 0.736 | 0.020 |
| 2444809.3323 | 0.8607 | -0.208 | 0.744 | 0.075 |
| 2444809.3351 | 0.8610 | -0.203 | 0.763 | 0.109 |
| 2444809.3386 | 0.8613 | -0.244 | 0.785 | 0.098 |
| 2444809.3427 | 0.8617 | -0.216 | 0.767 | 0.058 |
| 2444809.3455 | 0.8619 | -0.222 | 0.775 | 0.090 |
| 2444809.3483 | 0.8622 | -0.221 | 0.740 | 0.068 |
| 2444810.2865 | 0.9465 | - | 1.090 | 0.544 |
| 2444810.2913 | 0.9469 | 0.170 | 1.140 | 0.509 |
| 2444810.2948 | 0.9472 | 0.144 | 1.124 | 0.508 |
| 2444810.2983 | 0.9476 | 0.145 | 1.118 | 0.539 |
| 2444810.3094 | 0.9486 | 0.181 | 1.152 | 0.511 |
| 2444810.3129 | 0.9489 | 0.154 | 1.147 | 0.508 |
| 2444810.3177 | 0.9493 | 0.165 | 1.179 | 0.545 |
| 2444810.3226 | 0.9497 | 0.165 | 1.130 | 0.521 |
| 2444810.3344 | 0.9508 | 0.184 | 1.136 | 0.527 |
| 2444810.3372 | 0.9511 | 0.170 | 1.138 | 0.509 |
| 2444810.3406 | 0.9514 | 0.171 | 1.170 | 0.493 |
| 2444810.3434 | 0.9516 | - | 1.181 | 0.491 |
| 2444810.3462 | 0.9519 | 0.188 | 1.165 | 0.533 |
| 2444810.3497 | 0.9522 | 0.189 | 1.151 | 0.527 |
| 2444810.3531 | 0.9525 | 0.164 | 1.184 | 0.550 |
| 2444810.3559 | 0.9527 | 0.193 | 1.130 | 0.576 |
| 2444810.3601 | 0.9531 | 0.180 | 1.182 | 0.543 |
| 2444810.3649 | 0.9535 | 0.201 | 1.146 | 0.516 |
| 2444810.3677 | 0.9538 | 0.193 | 1.152 | 0.539 |
| 2444811.2733 | 0.0352 | 0.190 | 1.145 | 0.566 |
| 2444811.2767 | 0.0355 | 0.187 | 1.136 | 0.547 |
| 2444811.2795 | 0.0358 | 0.218 | 1.167 | 0.524 |
| 2444811.2830 | 0.0361 | 0.196 | 1.154 | 0.531 |
| 2444811.2868 | 0.0364 | 0.189 | 1.201 | 0.498 |
| 2444811.3017 | 0.0378 | 0.180 | 1.137 | 0.538 |
| 2444811.3045 | 0.0380 | 0.157 | 1.172 | 0.553 |
| 2444811.3073 | 0.0383 | 0.162 | 1.173 | 0.579 |
| 2444811.3108 | 0.0386 | 0.198 | 1.216 | 0.552 |
| 2444811.3136 | 0.0388 | 0.191 | 1.183 | 0.531 |
| 2444811.3163 | 0.0391 | 0.173 | 1.160 | 0.561 |
| 2444811.3191 | 0.0393 | 0.155 | 1.159 | 0.534 |
| 2444811.3337 | 0.0406 | 0.149 | 1.150 | 0.516 |
| 2444811.3365 | 0.0409 | 0.166 | 1.160 | 0.541 |
| 2444811.3392 | 0.0411 | 0.174 | 1.168 | 0.555 |
| 2444811.3434 | 0.0415 | 0.178 | 1.222 | 0.553 |
| 2444811.3462 | 0.0418 | 0.155 | 1.136 | 0.498 |
| 2444811.3504 | 0.0421 | 0.156 | 1.140 | 0.508 |
| 2444811.3531 | 0.0424 | 0.163 | 1.126 | 0.511 |
| 2444811.3573 | 0.0427 | 0.210 | 1.148 | 0.494 |
| 2444811.3601 | 0.0430 | 0.163 | 1.170 | 0.521 |
| 2444811.3656 | 0.0435 | 0.166 | 1.113 | 0.549 |
| 2444811.3684 | 0.0437 | 0.168 | 1.112 | 0.501 |
| 2444811.3719 | 0.0441 | 0.136 | 1.134 | 0.490 |
| 2444811.3747 | 0.0443 | 0.132 | 1.147 | 0.543 |
| 2444811.3781 | 0.0446 | 0.154 | 1.126 | 0.451 |
| 2444811.3809 | 0.0449 | 0.197 | 1.126 | 0.476 |
| 2444811.3844 | 0.0452 | 0.155 | - | 0.471 |
| 2444814.2693 | 0.3045 | -0.337 | 0.607 | -0.024 |
| 2444814.2721 | 0.3048 | -0.329 | 0.626 | -0.029 |
| 2444814.2742 | 0.3049 | -0.313 | 0.652 | 0.008 |
| 2444814.2776 | 0.3053 | -0.313 | 0.620 | -0.025 |
| 2444814.2860 | 0.3060 | -0.323 | 0.625 | -0.079 |
| 2444814.2888 | 0.3063 | -0.307 | 0.642 | -0.068 |
| 2444814.2943 | 0.3068 | -0.287 | 0.641 | -0.013 |
| 2444814.2971 | 0.3070 | -0.315 | 0.646 | -0.083 |
| 2444814.3006 | 0.3073 | -0.312 | 0.603 | -0.027 |
| 2444814.3033 | 0.3076 | -0.303 | 0.605 | -0.052 |
| 2444814.3061 | 0.3078 | -0.335 | 0.616 | -0.036 |
| 2444814.3103 | 0.3082 | -0.316 | 0.616 | -0.010 |
| 2444814.3131 | 0.3084 | -0.330 | 0.616 | -0.013 |
| 2444870.2512 | 0.3367 | -0.347 | 0.635 | -0.027 |
| 2444870.2547 | 0.3370 | -0.324 | 0.624 | -0.052 |
| 2444870.2609 | 0.3376 | -0.345 | 0.623 | -0.110 |
| 2444870.2644 | 0.3379 | -0.331 | 0.642 | -0.174 |
| 2444871.2415 | 0.4257 | -0.093 | 0.862 | 0.233 |
| 2444871.2443 | 0.4260 | -0.087 | 0.874 | 0.249 |
| 2444871.2470 | 0.4262 | -0.094 | 0.843 | 0.237 |
| 2444871.2498 | 0.4265 | -0.107 | 0.848 | 0.231 |
| 2444871.2533 | 0.4268 | -0.088 | 0.869 | 0.218 |
| 2444871.2561 | 0.4271 | -0.097 | 0.861 | 0.203 |
| 2444871.2588 | 0.4273 | -0.071 | 0.867 | 0.230 |
| 2444871.2616 | 0.4275 | -0.083 | 0.880 | 0.121 |
| 2444872.2422 | 0.5157 | 0.109 | 1.081 | 0.448 |
| 2444872.2449 | 0.5159 | 0.106 | 1.067 | 0.444 |
| 2444872.2477 | 0.5162 | 0.081 | 1.071 | 0.422 |
| 2444872.2526 | 0.5166 | 0.099 | 1.039 | 0.426 |
| 2444872.2575 | 0.5171 | 0.084 | 1.052 | 0.402 |
| 2444872.2602 | 0.5173 | 0.105 | 1.037 | 0.446 |
| 2444872.2630 | 0.5176 | 0.103 | 1.062 | 0.420 |
| 2444872.2665 | 0.5179 | 0.116 | 1.057 | 0.425 |
| 2444872.2699 | 0.5182 | 0.119 | 1.090 | 0.419 |
| 2444872.2727 | 0.5184 | 0.103 | 1.070 | 0.426 |
| 2444872.2762 | 0.5187 | 0.129 | 1.065 | 0.420 |
| 2444872.2790 | 0.5190 | 0.101 | 1.064 | 0.426 |
| 2444872.2824 | 0.5193 | 0.097 | 1.060 | 0.390 |
| 2444872.2852 | 0.5196 | 0.117 | 1.056 | 0.392 |
| 2444872.2887 | 0.5199 | 0.115 | 1.067 | 0.375 |
| 2444872.2915 | 0.5201 | 0.103 | 1.051 | 0.391 |
| 2444872.2943 | 0.5204 | 0.092 | 1.055 | 0.327 |
| 2445116.4264 | 0.4654 | 0.086 | 1.051 | 0.308 |
| 2445116.4291 | 0.4656 | 0.100 | 1.030 | 0.360 |
| 2445116.4319 | 0.4659 | 0.091 | 1.061 | 0.356 |
| 2445116.4347 | 0.4662 | 0.076 | 1.057 | 0.375 |
| 2445116.4382 | 0.4665 | 0.069 | 1.025 | 0.307 |
| 2445117.3866 | 0.5517 | 0.017 | 0.915 | 0.276 |
| 2445117.3907 | 0.5521 | 0.014 | 0.958 | 0.304 |
| 2445117.3956 | 0.5525 | 0.006 | 0.935 | 0.304 |
| 2445117.4006 | 0.5530 | -0.005 | 0.952 | 0.277 |
| 2445117.4060 | 0.5535 | -0.006 | 0.962 | 0.278 |
| 2445117.4101 | 0.5538 | -0.003 | 0.969 | 0.331 |
| 2445117.4144 | 0.5542 | -0.026 | 0.970 | 0.328 |
| 2445118.3905 | 0.6420 | -0.249 | 0.722 | 0.047 |
| 2445118.3933 | 0.6422 | -0.239 | 0.718 | 0.028 |
| 2445118.3961 | 0.6425 | -0.237 | 0.726 | 0.039 |
| 2445118.3989 | 0.6427 | -0.237 | 0.724 | 0.054 |
| 2445118.4010 | 0.6429 | -0.255 | 0.727 | 0.053 |
| 2445118.4037 | 0.6431 | -0.250 | 0.722 | 0.049 |
| 2445133.3073 | 0.9828 | 0.212 | 1.224 | 0.433 |
| 2445133.3108 | 0.9831 | 0.251 | 1.207 | 0.473 |
| 2445133.3143 | 0.9835 | 0.233 | 1.216 | 0.477 |
| 2445133.3177 | 0.9838 | 0.239 | 1.219 | 0.525 |
| 2445133.3205 | 0.9840 | 0.258 | 1.237 | 0.555 |
| 2445133.3233 | 0.9843 | 0.256 | 1.221 | 0.498 |
| 2445133.3281 | 0.9847 | 0.210 | 1.175 | 0.522 |
| 2445133.3330 | 0.9851 | 0.236 | 1.209 | 0.529 |
| 2445133.3358 | 0.9854 | 0.232 | 1.211 | 0.589 |
| 2445133.3386 | 0.9856 | 0.247 | 1.203 | 0.562 |
| 2445133.3403 | 0.9858 | 0.253 | 1.239 | 0.606 |
| 2445133.3441 | 0.9861 | 0.278 | 1.248 | 0.618 |
| 2445133.3476 | 0.9865 | 0.261 | 1.259 | 0.557 |
| 2445135.3490 | 0.1664 | -0.260 | 0.708 | 0.018 |
| 2445135.3504 | 0.1665 | -0.245 | 0.715 | 0.017 |
| 2445135.3518 | 0.1666 | -0.244 | 0.715 | 0.042 |
| 2445135.3559 | 0.1670 | -0.242 | 0.712 | 0.026 |
| 2445135.3601 | 0.1674 | -0.259 | 0.741 | 0.032 |
| 2445135.3622 | 0.1675 | -0.271 | 0.728 | 0.037 |
| 2445135.3649 | 0.1678 | -0.263 | 0.719 | 0.039 |
| 2445135.3684 | 0.1681 | -0.232 | 0.718 | 0.018 |
| 2445135.3712 | 0.1684 | -0.251 | 0.711 | 0.058 |
| 2445135.3740 | 0.1686 | -0.271 | 0.732 | 0.037 |
| 2445137.3587 | 0.3470 | -0.237 | 0.707 | -0.027 |
| 2445137.3649 | 0.3476 | -0.283 | 0.652 | 0.060 |
| 2445137.3705 | 0.3481 | -0.290 | 0.679 | -0.029 |
| 2445137.3740 | 0.3484 | -0.308 | 0.655 | -0.029 |
| 2445137.3768 | 0.3486 | -0.241 | 0.654 | -0.012 |
| 2445137.3795 | 0.3489 | -0.294 | 0.664 | -0.036 |
| 2445137.3830 | 0.3492 | -0.282 | 0.658 | -0.004 |
| 2445140.3773 | 0.6184 | -0.179 | 0.760 | 0.028 |
| 2445140.3801 | 0.6186 | -0.182 | 0.769 | 0.106 |
| 2445140.3829 | 0.6189 | -0.185 | 0.782 | 0.116 |
| 2445140.3857 | 0.6191 | -0.101 | 0.764 | 0.161 |
| 2445140.3885 | 0.6194 | -0.180 | 0.764 | 0.106 |
| 2445140.3912 | 0.6196 | -0.179 | 0.763 | 0.181 |
| 2445140.3940 | 0.6199 | -0.182 | 0.830 | 0.145 |
| 2445140.3989 | 0.6203 | -0.181 | 0.829 | 0.164 |
| 2445144.3489 | 0.9754 | 0.180 | 1.159 | 0.485 |
| 2445144.3517 | 0.9756 | 0.190 | 1.175 | 0.489 |
| 2445144.3544 | 0.9759 | 0.183 | 1.169 | 0.521 |
| 2445144.3565 | 0.9760 | 0.200 | 1.206 | 0.529 |
| 2445144.3593 | 0.9763 | 0.209 | 1.198 | 0.504 |
| 2445144.3621 | 0.9765 | 0.297 | 1.186 | 0.491 |
| 2445144.3635 | 0.9767 | 0.203 | 1.201 | 0.487 |
| 2445144.3683 | 0.9771 | 0.197 | 1.183 | 0.548 |
| 2445144.3711 | 0.9774 | 0.211 | 1.207 | 0.531 |
| 2445144.3746 | 0.9777 | 0.202 | 1.170 | 0.464 |
| 2445144.3773 | 0.9779 | 0.201 | 1.214 | 0.498 |
| 2445144.3801 | 0.9782 | 0.182 | 1.195 | 0.491 |
| 2445144.3822 | 0.9784 | 0.182 | 1.183 | 0.520 |
| 2445144.3850 | 0.9786 | 0.199 | 1.225 | 0.523 |
| 2445144.3885 | 0.9789 | 0.188 | 1.208 | 0.517 |
| 2445144.3947 | 0.9795 | 0.210 | 1.182 | 0.455 |
| 2445144.3968 | 0.9797 | 0.200 | 1.175 | 0.483 |
| 2445144.3992 | 0.9799 | 0.196 | 1.163 | 0.485 |
| 2445145.3364 | 0.0641 | 0.095 | 1.085 | 0.397 |
| 2445145.3392 | 0.0644 | 0.114 | 1.044 | 0.391 |
| 2445145.3419 | 0.0646 | 0.098 | 1.068 | 0.417 |
| 2445145.3461 | 0.0650 | 0.082 | 1.060 | 0.377 |
| 2445145.3489 | 0.0652 | 0.085 | 1.072 | 0.406 |
| 2445145.3517 | 0.0655 | 0.081 | 1.081 | 0.392 |
| 2445145.3554 | 0.0658 | 0.081 | 1.067 | 0.368 |
| 2445145.3572 | 0.0660 | 0.101 | 1.054 | 0.365 |
| 2445145.3600 | 0.0662 | 0.093 | 1.087 | 0.419 |
| 2445145.3621 | 0.0664 | 0.106 | 1.063 | 0.449 |
| 2445145.3648 | 0.0667 | - | 1.064 | 0.404 |
| 2445145.3672 | 0.0669 | 0.110 | 1.034 | 0.443 |
| 2445145.3704 | 0.0672 | 0.082 | 1.029 | 0.398 |
| 2445145.3732 | 0.0674 | 0.106 | 1.093 | 0.416 |
| 2445145.3767 | 0.0677 | 0.068 | 1.036 | 0.430 |
| 2445150.4391 | 0.5228 | 0.094 | 1.015 | 0.343 |
| 2445150.4418 | 0.5231 | 0.075 | 1.028 | 0.383 |
| 2445150.4446 | 0.5233 | 0.099 | 1.021 | 0.336 |
| 2445150.4467 | 0.5235 | 0.070 | 1.041 | 0.365 |
| 2445150.4502 | 0.5238 | 0.080 | 1.028 | 0.318 |
| 2445150.4530 | 0.5241 | 0.075 | 1.046 | 0.315 |
| 2445150.4550 | 0.5242 | 0.074 | 1.009 | 0.301 |
| 2445150.4578 | 0.5245 | 0.062 | 1.037 | 0.337 |
| 2445150.4599 | 0.5247 | 0.063 | 1.023 | 0.309 |
| 2445150.4627 | 0.5249 | 0.063 | 1.012 | 0.319 |
| 2445176.3254 | 0.8497 | -0.270 | 0.688 | 0.050 |
| 2445176.3281 | 0.8500 | -0.238 | 0.703 | -0.011 |
| 2445176.3309 | 0.8502 | -0.259 | 0.717 | 0.037 |
| 2445176.3337 | 0.8505 | -0.262 | 0.703 | 0.116 |
| 2445176.3365 | 0.8507 | -0.247 | 0.713 | 0.030 |
| 2445176.3386 | 0.8509 | -0.252 | 0.716 | 0.043 |
| 2445176.3413 | 0.8512 | -0.266 | 0.703 | 0.048 |
| 2445176.3441 | 0.8514 | -0.249 | 0.719 | 0.039 |
| 2445176.3504 | 0.8520 | -0.264 | 0.732 | 0.044 |
| 2445176.3531 | 0.8522 | -0.263 | 0.680 | -0.003 |
| 2445196.3021 | 0.6454 | -0.251 | 0.703 | 0.028 |
| 2445196.3062 | 0.6458 | -0.252 | 0.681 | 0.045 |
| 2445196.3104 | 0.6462 | -0.249 | 0.694 | 0.037 |
| 2445196.3156 | 0.6466 | -0.248 | 0.699 | 0.020 |
| 2445196.3187 | 0.6469 | -0.264 | 0.717 | -0.006 |
| 2445196.3271 | 0.6477 | -0.264 | 0.707 | -0.003 |
| 2445196.3319 | 0.6481 | -0.271 | 0.693 | -0.029 |
| 2445196.3382 | 0.6487 | -0.238 | 0.703 | -0.018 |
| 2445196.3417 | 0.6490 | -0.278 | 0.680 | -0.065 |
| 2445198.2565 | 0.8211 | -0.291 | 0.667 | 0.032 |
| 2445198.2655 | 0.8219 | -0.301 | 0.653 | -0.002 |
| 2445198.2703 | 0.8224 | -0.308 | 0.665 | 0.026 |
| 2445198.2752 | 0.8228 | -0.302 | 0.668 | 0.021 |
| 2445198.2821 | 0.8234 | -0.281 | 0.656 | 0.005 |
| 2445198.2870 | 0.8239 | -0.298 | 0.645 | 0.035 |
| 2445198.2905 | 0.8242 | -0.289 | 0.659 | 0.014 |
| 2445198.2974 | 0.8248 | -0.289 | 0.661 | 0.008 |
| 2445198.3016 | 0.8252 | -0.279 | 0.640 | 0.004 |
| 2445198.3057 | 0.8255 | -0.294 | 0.669 | -0.027 |
| 2445199.2621 | 0.9115 | -0.036 | 0.891 | 0.210 |
| 2445199.2677 | 0.9120 | -0.033 | 0.937 | 0.259 |
| 2445199.2718 | 0.9124 | -0.031 | 0.950 | 0.222 |
| 2445199.2822 | 0.9133 | -0.025 | 0.930 | 0.242 |
| 2445199.2864 | 0.9137 | -0.074 | 0.982 | 0.202 |
| 2445199.2909 | 0.9141 | -0.036 | 0.882 | 0.223 |
| 2445199.2933 | 0.9143 | -0.044 | 0.911 | 0.185 |
| 2445199.2961 | 0.9146 | -0.053 | 0.891 | 0.159 |
| 2445231.2226 | 0.7844 | -0.320 | 0.647 | -0.045 |
| 2445231.2254 | 0.7847 | -0.330 | 0.641 | -0.038 |
| 2445231.2282 | 0.7849 | -0.313 | 0.661 | -0.019 |
| 2445231.2310 | 0.7852 | -0.319 | 0.658 | -0.008 |
| 2445231.2331 | 0.7854 | -0.318 | 0.659 | -0.007 |
| 2445231.2358 | 0.7856 | -0.337 | 0.640 | 0.026 |
| 2445231.2386 | 0.7859 | -0.320 | 0.631 | -0.030 |
| 2445231.2414 | 0.7861 | -0.330 | 0.665 | -0.032 |
| 2445231.2435 | 0.7863 | -0.317 | 0.622 | -0.059 |
| 2445232.2061 | 0.8728 | -0.212 | 0.757 | 0.103 |
| 2445232.2088 | 0.8731 | -0.206 | 0.759 | 0.118 |
| 2445232.2109 | 0.8733 | -0.212 | 0.759 | 0.126 |
| 2445232.2137 | 0.8735 | -0.196 | 0.760 | 0.103 |
| 2445232.2158 | 0.8737 | -0.209 | 0.750 | 0.081 |
| 2445232.2186 | 0.8740 | -0.213 | 0.767 | 0.067 |
| 2445232.2202 | 0.8741 | -0.213 | 0.759 | 0.116 |
| 2445232.2234 | 0.8744 | -0.210 | 0.754 | 0.120 |
| 2445232.2255 | 0.8746 | -0.201 | 0.764 | 0.077 |
| 2445232.2283 | 0.8748 | -0.212 | 0.731 | 0.088 |
| 2445259.1754 | 0.2971 | -0.346 | 0.607 | -0.052 |
| 2445259.1775 | 0.2973 | -0.360 | 0.634 | -0.052 |
| 2445259.1803 | 0.2976 | -0.366 | 0.632 | -0.058 |
| 2445259.1831 | 0.2978 | -0.386 | 0.636 | -0.046 |
| 2445259.1851 | 0.2980 | -0.354 | 0.634 | -0.095 |
| 2445259.1879 | 0.2982 | -0.355 | 0.613 | -0.139 |
| 2445259.1914 | 0.2985 | - | 0.619 | -0.113 |
| 2445260.1501 | 0.3847 | - | 0.741 | 0.084 |
| 2445260.1622 | 0.3858 | -0.207 | 0.735 | 0.080 |
| 2445260.1647 | 0.3860 | -0.231 | 0.741 | 0.029 |
| 2445260.1685 | 0.3864 | -0.224 | 0.762 | 0.082 |
| 2445260.1706 | 0.3866 | -0.208 | 0.759 | 0.046 |
| 2445260.1733 | 0.3868 | -0.223 | 0.741 | 0.089 |
| 2445260.1761 | 0.3871 | - | 0.737 | 0.079 |
| 2445527.3278 | 0.4013 | -0.191 | 0.816 | 0.206 |
| 2445527.3327 | 0.4018 | -0.181 | 0.841 | 0.132 |
| 2445527.3369 | 0.4021 | -0.212 | 0.730 | 0.176 |
| 2445527.3410 | 0.4025 | -0.197 | 0.771 | 0.169 |
| 2445527.3445 | 0.4028 | -0.177 | 0.746 | 0.171 |
| 2445527.3487 | 0.4032 | -0.176 | 0.740 | 0.065 |
| 2445527.3521 | 0.4035 | -0.188 | 0.735 | 0.175 |
| 2445528.3535 | 0.4935 | 0.143 | 1.053 | 0.459 |
| 2445528.3570 | 0.4938 | 0.126 | 1.106 | 0.401 |
| 2445528.3598 | 0.4941 | 0.103 | 1.049 | 0.392 |
| 2445528.3653 | 0.4946 | 0.106 | 1.065 | 0.411 |
| 2445528.3681 | 0.4948 | 0.104 | 1.052 | 0.399 |
| 2445528.3709 | 0.4951 | 0.106 | 1.071 | 0.361 |
| 2445528.3737 | 0.4953 | 0.117 | 1.063 | 0.385 |
| 2445579.2287 | 0.0667 | 0.065 | 1.042 | 0.373 |
| 2445579.2314 | 0.0669 | 0.058 | 1.029 | 0.353 |
| 2445579.2342 | 0.0672 | 0.040 | 1.017 | 0.367 |
| 2445579.2377 | 0.0675 | 0.064 | 1.015 | 0.395 |
| 2445579.2405 | 0.0678 | 0.073 | 1.047 | 0.382 |
| 2445579.2432 | 0.0680 | 0.060 | 0.996 | 0.343 |
| 2445579.2460 | 0.0683 | 0.062 | 1.030 | 0.353 |
| 2445579.2488 | 0.0685 | 0.018 | 0.980 | 0.309 |
| 2445616.1905 | 0.3892 | -0.101 | 0.840 | 0.194 |
| 2445616.1953 | 0.3896 | -0.191 | 0.758 | 0.134 |
| 2445616.1988 | 0.3899 | -0.123 | 0.871 | 0.172 |
| 2445616.2023 | 0.3903 | -0.116 | 0.855 | 0.164 |
| 2445616.2051 | 0.3905 | -0.105 | 0.856 | 0.163 |
| 2445616.2085 | 0.3908 | -0.064 | 0.826 | 0.189 |
| 2445616.2155 | 0.3914 | -0.118 | 0.841 | 0.189 |
| 2445617.1795 | 0.4781 | 0.101 | 1.105 | 0.456 |
| 2445617.1829 | 0.4784 | 0.101 | 1.086 | 0.475 |
| 2445617.1850 | 0.4786 | 0.121 | 1.080 | 0.472 |
| 2445617.1878 | 0.4788 | 0.115 | 1.089 | 0.447 |
| 2445617.1906 | 0.4791 | 0.111 | 1.099 | 0.402 |
| 2445617.1934 | 0.4793 | 0.107 | 1.104 | 0.442 |
| 2445617.1961 | 0.4796 | 0.140 | 1.093 | 0.434 |
| 2445875.3539 | 0.6854 | -0.246 | 0.759 | 0.051 |
| 2445875.3577 | 0.6857 | -0.237 | 0.758 | 0.052 |
| 2445875.3605 | 0.6860 | -0.255 | 0.751 | 0.111 |
| 2445875.3636 | 0.6863 | -0.244 | 0.744 | 0.081 |
| 2445875.3666 | 0.6865 | -0.227 | 0.727 | 0.099 |
| 2445875.3697 | 0.6868 | -0.227 | 0.742 | 0.086 |
| 2445875.3735 | 0.6871 | -0.237 | 0.744 | 0.078 |
| 2445875.3763 | 0.6874 | -0.237 | 0.761 | 0.092 |
| 2445886.3659 | 0.6753 | -0.235 | 0.772 | 0.020 |
| 2445886.3690 | 0.6755 | -0.228 | 0.775 | 0.040 |
| 2445886.3725 | 0.6758 | -0.210 | 0.789 | 0.050 |
| 2445886.3757 | 0.6761 | -0.208 | 0.787 | 0.070 |
| 2445886.3806 | 0.6766 | -0.194 | 0.763 | 0.132 |
| 2445886.3839 | 0.6769 | -0.203 | 0.748 | 0.080 |
| 2445886.3921 | 0.6776 | -0.239 | 0.758 | 0.118 |
| 2445886.3954 | 0.6779 | -0.228 | 0.756 | 0.068 |
| 2445886.3986 | 0.6782 | -0.225 | 0.779 | 0.079 |
| 2445886.4020 | 0.6785 | -0.280 | 0.686 | 0.025 |
| 2445886.4053 | 0.6788 | -0.241 | 0.746 | 0.034 |
| 2445915.3076 | 0.2768 | -0.211 | 0.765 | 0.045 |
| 2445915.3116 | 0.2772 | -0.191 | 0.787 | 0.103 |
| 2445915.3148 | 0.2775 | -0.208 | 0.763 | 0.107 |
| 2445915.3195 | 0.2779 | -0.217 | 0.766 | 0.085 |
| 2445915.3227 | 0.2782 | -0.223 | 0.751 | 0.053 |
| 2445915.3260 | 0.2785 | -0.227 | 0.753 | 0.101 |
| 2445915.3293 | 0.2788 | -0.225 | 0.748 | 0.022 |
| 2445968.2046 | 0.0317 | 0.187 | 1.174 | 0.554 |
| 2445968.2094 | 0.0322 | 0.161 | 1.150 | 0.553 |
| 2445968.2137 | 0.0326 | 0.191 | 1.165 | 0.619 |
| 2445968.2178 | 0.0329 | 0.186 | 1.161 | 0.511 |
| 2445968.2211 | 0.0332 | 0.154 | 1.166 | 0.523 |
| 2445968.2304 | 0.0341 | 0.169 | 1.192 | 0.526 |
| 2445968.2346 | 0.0344 | 0.150 | 1.128 | 0.483 |
| 2445968.2411 | 0.0350 | 0.159 | 1.138 | 0.465 |
| 2445968.2447 | 0.0353 | 0.150 | 1.150 | 0.427 |
| 2445968.2484 | 0.0357 | 0.180 | 1.196 | 0.451 |
| 2445968.2533 | 0.0361 | 0.152 | 1.140 | 0.381 |
| 2445875.3539 | 0.6854 | -0.246 | 0.759 | 0.051 |
| 2445875.3577 | 0.6857 | -0.237 | 0.758 | 0.052 |
| 2445875.3605 | 0.6860 | -0.255 | 0.751 | 0.111 |
| 2445875.3636 | 0.6863 | -0.244 | 0.744 | 0.081 |
| 2445875.3666 | 0.6865 | -0.227 | 0.727 | 0.099 |
| 2445875.3697 | 0.6868 | -0.227 | 0.742 | 0.080 |
| 2445875.3735 | 0.6871 | -0.237 | 0.744 | 0.078 |
| 2445875.3763 | 0.6874 | -0.237 | 0.761 | 0.092 |
| 2445886.3659 | 0.6753 | -0.235 | 0.772 | 0.020 |
| 2445886.3690 | 0.6755 | -0.228 | 0.775 | 0.040 |
| 2445886.3725 | 0.6758 | -0.210 | 0.789 | 0.050 |
| 2445886.3757 | 0.6761 | -0.208 | 0.787 | 0.070 |
| 2445886.3806 | 0.6766 | -0.194 | 0.763 | 0.132 |
| 2445886.3839 | 0.6769 | -0.203 | 0.748 | 0.080 |
| 2445886.3921 | 0.6776 | -0.239 | 0.758 | 0.118 |
| 2445886.3954 | 0.6779 | -0.228 | 0.756 | 0.068 |
| 2445886.3986 | 0.6782 | -0.225 | 0.779 | 0.079 |
| 2445886.4020 | 0.6785 | -0.280 | 0.686 | 0.025 |
| 2445886.4053 | 0.6788 | -0.241 | 0.746 | 0.034 |
| 2445915.3076 | 0.2768 | -0.211 | 0.765 | 0.045 |
| 2445915.3116 | 0.2772 | -0.191 | 0.787 | 0.103 |
| 2445915.3148 | 0.2775 | -0.208 | 0.763 | 0.107 |
| 2445915.3195 | 0.2779 | -0.217 | 0.766 | 0.085 |
| 2445915.3227 | 0.2782 | -0.223 | 0.751 | 0.053 |
| 2445915.3260 | 0.2785 | -0.227 | 0.753 | 0.101 |
| 2445915.3243 | 0.2783 | -0.225 | 0.748 | 0.022 |
| 2445968.2046 | 0.0317 | 0.187 | 1.174 | 0.554 |
| 2445968.2094 | 0.0322 | 0.161 | 1.150 | 0.553 |
| 2445968.2137 | 0.0326 | 0.191 | 1.165 | 0.619 |
| 2445968.2178 | 0.0329 | 0.186 | 1.161 | 0.511 |
| 2445968.2211 | 0.0332 | 0.154 | 1.166 | 0.523 |
| 2445968.2304 | 0.0341 | 0.169 | 1.192 | 0.526 |
| 2445968.2346 | 0.0344 | 0.150 | 1.128 | 0.483 |
| 2445968.2411 | 0.0350 | 0.159 | 1.138 | 0.465 |
| 2445968.2447 | 0.0353 | 0.150 | 1.150 | 0.427 |
| 2445968.2484 | 0.0357 | 0.180 | 1.196 | 0.451 |
| 2445968.2533 | 0.0361 | 0.152 | 1.134 | 0.381 |
| 2446226.4194 | 0.2427 | -0.279 | 0.670 | -0.003 |
| 2446226.4241 | 0.2431 | -0.254 | 0.658 | -0.054 |
| 2446226.4275 | 0.2434 | -0.268 | 0.683 | -0.035 |
| 2446226.4307 | 0.2437 | -0.282 | 0.678 | -0.014 |
| 2446226.4341 | 0.2440 | -0.272 | 0.693 | 0.028 |
| 2446227.4285 | 0.3334 | -0.216 | 0.759 | 0.056 |
| 2446227.4321 | 0.3337 | -0.221 | 0.724 | -0.083 |
| 2446227.4358 | 0.3340 | -0.221 | 0.718 | -0.076 |
| 2446227.4391 | 0.3343 | -0.235 | 0.746 | 0.068 |
| 2446227.4430 | 0.3347 | -0.236 | 0.731 | - |
| 2446238.3664 | 0.3166 | -0.309 | 0.679 | -0.043 |
| 2446238.3690 | 0.3168 | -0.321 | 0.670 | 0.009 |
| 2446238.3715 | 0.3170 | -0.313 | 0.668 | -0.008 |
| 2446238.3741 | 0.3173 | -0.299 | 0.658 | 0.000 |
| 2446238.3767 | 0.3175 | -0.317 | 0.667 | 0.016 |
| 2446258.3489 | 0.1128 | -0.090 | 0.895 | 0.117 |
| 2446258.3514 | 0.1130 | -0.102 | 0.874 | 0.192 |
| 2446258.3540 | 0.1133 | -0.107 | 0.851 | 0.170 |
| 2446258.3567 | 0.1135 | -0.055 | 0.879 | 0.199 |
| 2446258.3595 | 0.1138 | -0.105 | 0.830 | 0.136 |
| 2446258.3621 | 0.1140 | -0.118 | 0.842 | 0.169 |
| 2446286.2626 | 0.6220 | -0.207 | 0.744 | 0.140 |
| 2446286.2651 | 0.6222 | -0.212 | 0.705 | 0.078 |
| 2446286.2675 | 0.6224 | -0.230 | 0.720 | 0.122 |
| 2446286.2701 | 0.6226 | -0.206 | 0.735 | - |
| 2446286.2723 | 0.6228 | -0.185 | 0.786 | 0.188 |
| 2446317.2719 | 0.4094 | -0.221 | 0.732 | -0.002 |
| 2446317.2745 | 0.4096 | -0.193 | 0.789 | 0.101 |
| 2446317.2771 | 0.4099 | -0.186 | 0.782 | 0.092 |
| 2446317.2798 | 0.4101 | -0.207 | 0.770 | 0.024 |
| 2446317.2822 | 0.4103 | -0.200 | 0.799 | 0.029 |
| 2446317.2847 | 0.4105 | -0.196 | 0.810 | 0.028 |
| 2446317.2872 | 0.4108 | -0.201 | 0.794 | 0.101 |